\documentclass{cernrep}
 \usepackage{ifpdf}
 \ifpdf
 \usepackage[pdftex]{hyperref}
 \else
 \usepackage[hypertex]{hyperref}
 \fi

 \hypersetup{
   pdftitle={},%
   pdfauthor={},%
   pdfsubject={},%
   pdfkeywords={},%
   pdfstartview={},%
   bookmarksopen=true, breaklinks=true, debug=true, %
   colorlinks=true, linkcolor=red, citecolor=blue, urlcolor=blue
 }

\begin{document}
\title{Quarkonium-photoproduction prospects at a fixed-target experiment at the LHC (AFTER@LHC)}
\author{L. Massacrier$^{1}$, J.P. Lansberg$^{1}$, L. Szymanowski$^{2}$ and J. Wagner$^{2}$}
\institute{
1. IPNO, Univ. Paris-Sud, CNRS/IN2P3, Universit\'e Paris-Saclay, 91406 Orsay, France.\\
2. National Centre for Nuclear Research (NCBJ), Hoza 69,00-681, Warsaw,Poland. }

\begin{abstract}
We report on the potentialities offered by a fixed-target experiment at the LHC using the proton and ion LHC beams (AFTER@LHC project) regarding the study of $J/\psi$ exclusive-photoproduction in $pA$ and $AA$ collisions. The foreseen usage of polarised targets (hydrogen, deuteron, helium) gives access to measurements of the Single-Transverse-Spin Asymmetries of this exclusive process, therefore allowing one to access the helicity-flip Generalised Parton Distribution (GPD) $E_{g}$. We detail the expected yields of photoproduced $J/\psi$ in proton-hydrogen and lead-hydrogen collisions and discuss the statistical uncertainties on the asymmetry measurement for one year of data taking at the LHC. 

\end{abstract}

\keywords{AFTER@LHC, fixed-target experiment, photoproduction, $J/\psi$, GPD, Single Transverse Spin Asymmetries}

\maketitle

\section{The AFTER@LHC project}

AFTER@LHC is a proposal for a multi-purpose fixed-target programme using the multi-TeV proton or heavy ion beams of the LHC to perform the most energetic fixed-target collisions ever performed so far \cite{AFTERweb}. If used in a fixed-target mode, the LHC beams offer the possibility to study with high precision pH or $pA$ collisions at $\sqrt{s_{NN}}$ = 115 GeV and PbA collisions at $\sqrt{s_{NN}}$ = 72 GeV, where $A$ is the atomic mass of the target. The fixed-target mode offers several unique assets compared to the collider mode: outstanding luminosities can be reached thanks to the high density of the target; standard detectors can access the far backward center-of-mass (c.m) region thanks to the boost between the colliding-nucleon c.m system and the laboratory system (this region remains completely uncharted with hard reactions); the target species can easily be changed; polarised target can be used for spin studies. The physics opportunities offered by a fixed-target programme at LHC have been published in \cite{Brodsky:2012vg, Lansberg:2016urh, Lansberg:2012kf, Lansberg:2012wj, Rakotozafindrabe:2012ei, Lorce:2012rn, Lansberg:2012sq, Rakotozafindrabe:2013au, Lansberg:2013wpx, Rakotozafindrabe:2013cmt, Lansberg:2014myg, Massacrier:2015nsm, Massacrier:2015qba, Lansberg:2016gwm, Kikola:2017hnp, Trzeciak:2017csa} and are summarised below in three main objectives. 

First, whereas the need for precise measurements of the partonic structure of nucleons and nuclei at small momentum fraction $x$ is usually highlighted as a strong motivation for new large-scale experimental facilities, the structure of nucleon and nuclei at high $x$ is probably as poorly known with long-standing puzzles such as the EMC effect \cite{Aubert:1983xm} in the nuclei or a possible non-perturbative source of charm or beauty quarks in the proton which would carry a relevant fraction of its momentum \cite{Brodsky:1980pb}. With an extensive coverage of the backward region corresponding to high $x$ in the target, AFTER@LHC is very well placed for performing this physics with a hadron beam. 

The second objective of AFTER@LHC is the search and characterisation of the Quark-Gluon Plasma (QGP), a deconfined state of nuclear matter, which was prevailing in the universe few microseconds after the Big Bang. QGP is expected to be formed when the surrounding hadronic matter is extremely compressed or heated. These conditions can be achieved in ultra-relativistic Heavy-Ion collisions (HI). AFTER@LHC with a c.m. energy of 72 GeV provides a complementary coverage to the RHIC- and SPS- based experiments, in the region of high temperatures and low baryon-chemical potentials, where the QGP is expected to be produced. AFTER@LHC will provide crucial information about the phase transition by: (i) scanning the longitudinal extension of the hot medium, (ii) colliding systems of different sizes, (iii) analysing the centrality dependence of these collisions. Together they should provide a measurement of the temperature dependence of the system viscosity both as a QGP or a hadron gas. Additionally, measurements of production of various quarkonia states in HI collisions can provide insight into thermodynamic properties of the QGP. Their sequential suppression was predicted to occur in the deconfined partonic medium due to Debye screening of the quark-antiquark potential \cite{Matsui:1986dk}. However, other effects (Cold Nuclear Matter effects (CNM), feed-down, secondary production via coalescence...) can also alter the observed yields. AFTER@LHC will provide a complete set of quarkonia measurements (together with open heavy flavours) allowing one to access the temperature of the formed medium in $AA$ collisions, and cold nuclear matter effects in $pA$ ($AA$) collisions. Thanks to the large statistics expected, a golden probe will be the measurement of $\Upsilon$(nS) production in $pp$, $pA$ and $AA$ collisions, allowing one to calibrate the quarkonium thermometer and to search for the phase transition by looking at $\Upsilon$(nS) suppression (and other observables like charged hadron $v_{2}$) as a function of rapidity and the system size. 

Finally, despite decades of efforts, the internal structure of the nucleon and the distribution and dynamics of its constituents are still largely unknown. One of the most significant issues is our limited understanding of the spin structure of the nucleon, especially how its elementary constituents (quarks and gluons) bind into a spin-half object. Among others, Single Transverse Spin Asymmetries (STSA) in different hard-scattering processes are powerful observables to further explore the dynamics of the partons inside hadrons \cite{DAlesio:2007bjf}. 
Thanks to the large yields expected with AFTER@LHC, STSA of heavy-flavours and quarkonia --which are currently poorly known-- could be measured with high accuracy, if a polarised target can be installed \cite{Kikola:2017hnp}. 

We show here that AFTER@LHC can also rely
on quarkonium exclusive-photoproduction processes to explore the three-dimensional tomography of hadrons via Generalised Parton Distributions (GPDs) \cite{Diehl:2003ny}. Photoproduction is indeed accessible in Ultra-Peripheral Collisions (UPCs) and the quarkonium mass presumably sets the hard scale to use collinear factorisation in terms of (gluon) GPDs, which are directly related to the total angular momentum carried by quarks and gluons. In fact, exclusive $J/\psi$ production \cite{Ivanov:2004vd} drew much attention in the recent years due to the fact that it is sensitive, even at leading order, to gluon GPDs. Beside, with the addition of a polarised hydrogen\footnote{Measurements with deuteron and helium targets are also considered. } target, AFTER@LHC opens a unique possibility to study STSA of such process, which is sensitive to yet unknown GPD $E_{g}$ \cite{Koempel:2011rc}. In this contribution, we report on such studies through the collisions of proton and lead beams onto a polarised hydrogen target at AFTER@LHC energies. 

\section{Possible technical implementations at the LHC and projected luminosities}

Several promising technical solutions exist in order to perform fixed-target collisions at the LHC. One can either use an internal (solid or gaseous) target coupled to an already existing LHC detector or build a new beam line together with a new detector. The first solution can be achieved in a shorter time scale, at limited cost and civil engineering. Moreover the fixed-target programme can be simultaneously run  with the current LHC collider experiments, without affecting the LHC performances.

The direct injection of noble gases in the LHC beam pipe is currently being used by the LHCb collaboration with the SMOG device \cite{Aaij:2014ida}. However, this system has some limitations, in particular: (i) the gas density achieved inside the Vertex Locator of LHCb is small (of the order of $10^{-7}$ mbar); (ii) there is no possibility to inject polarised gas; (iii) there is no dedicated pumping system close to the target; (iv) the data taking time has so far been limited to few days in a row. The use of more complex systems with higher density gaseous and polarised target inside one of the existing LHC experiment is under study. For instance, an hydrogen gas jet is currently used at RHIC to measure the proton beam polarisation \cite{Zelenski:2005mz}.  The H-jet system consists of a polarised free atomic beam source cooled down to 80K, providing an hydrogen inlet flux of 1.3 $\times 10^{7}$ H/s. With such a device, the gas density can be increased by about one order of magnitude with respect to the SMOG device and probably be continuously run. Another promising alternative solution is the use of an openable storage cell placed inside the LHC beam pipe. Such a system was developed for the HERMES experiment \cite{Barschel:2015mka, Steffens:2015kvp}. Polarised hydrogen, deuteron and Helium-3 at densities about 200 times larger than the ones of the H-jet system can be injected, as well as heavier unpolarised gases. 

Fixed-target collisions can also be obtained with a solid target (wire, foil) interacting with the LHC beams. There are two ways of doing so: either thanks to a system which permit to move directly the target inside the beam halo \cite{Kurepin:2011zz}; or by using a bent crystal (see work by UA9 collaboration \cite{Scandale:2010zzb}) upstream of an existing experiment ($\sim$ 100 m) to deflect the beam halo onto the fixed target. In both cases, the target (or an assembly of several targets) can be placed at a few centimetres from the nominal interaction point, allowing one to fully exploit the performances of an existing detector. The usage of the bent crystal offers the additional advantage to better control the flux\footnote{The deflected halo beam flux is considered to be about 5 $\times$ 10$^{8}$ p/s and  10$^{5}$ Pb/s.} of particles sent on the target, and therefore the luminosity determination. Most probably such simple solid targets could not be polarised. The spin physics part of the AFTER@LHC programme could naturally not be conducted with such an option. 

Tab. \ref{Tab1} summarises the target areal density, the beam flux intercepting the target, the expected instantaneous and yearly integrated luminosities for $p$H and PbH collisions, for the possible technical solutions described above. Luminosities as large as 10 fb$^{-1}$ can be collected in $p$H collisions in one LHC year with a storage cell. In PbH collisions, similar luminosities ($\sim$ 100 nb$^{-1}$) can be reached with a gas-jet or a storage cell since the gas density has to be levelled in order to avoid a too large beam consumption\footnote{Assuming a total cross section $ \sigma_{\rm PbH}$ = 3 barn, 15$\%$ of the beam is used over a fill.}. We stress that these are annual numbers and they can be cumulated over different runs.

\begin{table}[!htpb]
\caption{Target areal density $\theta_{\rm target}$, beam flux intercepting the target, expected instantaneous and yearly integrated luminosities for pH and PbH collisions, for the possible technical solutions described in the text. The solid target is considered to be 5 mm thick along the beam direction. The LHC year is considered to last 10$^{7}$s for the proton beam and 10$^{6}$s for the lead beam.}
\resizebox{\textwidth}{!}{%
\begin{tabular}{|p{4.5cm}|c|c|c|c|c|c|c|c|}
\hline
 & \multicolumn{2}{c|}{$\theta_{target}$ (cm$^{-2}$)} &  \multicolumn{2}{c|}{Beam flux (s$^{-1}$)} & \multicolumn{2}{c|}{$\cal{L}$ (cm$^{-2}$.s$^{-1}$)} & \multicolumn{2}{c|}{$\cal{L}_{\rm int}$}  \\
  & \multicolumn{2}{c|}{} &  \multicolumn{2}{c|}{} & \multicolumn{2}{c|}{} & (fb$^{-1}$) & (nb$^{-1}$)  \\
\hline
Technical solutions & pH  & PbH & pH & PbH & pH & Pb & pH & PbH \\
\hline
 Gas-jet    & 1.2 $\times$ 10$^{12}$ & 2.54 $\times$ 10$^{14}$  & 3.63 $\times$ 10$^{18}$  & 4.66 $\times$ 10$^{14}$ & 4.4 $\times$ 10$^{30}$ & 1.2 $\times$ 10$^{29}$  & $\sim$ 0.05 & $\sim$ 100 \\
 \hline
Storage Cell  & 2.5 $\times$ 10$^{14}$ & 2.54 $\times$ 10$^{14}$  & 3.63 $\times$ 10$^{18}$  & 4.66 $ \times$ 10$^{14}$  &  9.1 $\times$ 10$^{32}$ & 1.2 $\times$ 10$^{29}$  & $\sim$ 10 & $\sim$ 100  \\
\hline
Bent Crystal + Solid target    & 2.6 $\times$ 10$^{22}$ & 2.6 $\times$ 10$^{22}$  & 5 $\times$ 10$^{8}$  & 10$^{5}$   & 1.3 $\times$ 10$^{31}$ & 2.6 $\times$ 10$^{27}$ & $\sim$ 0.15 & $\sim$ 3\\
\hline
\end{tabular} }
\label{Tab1}
\end{table}

\section{Prospects for $J/\psi$ photoproduction studies with AFTER@LHC}

\begin{figure*}
\centering
 \includegraphics[width=7.9cm,clip]{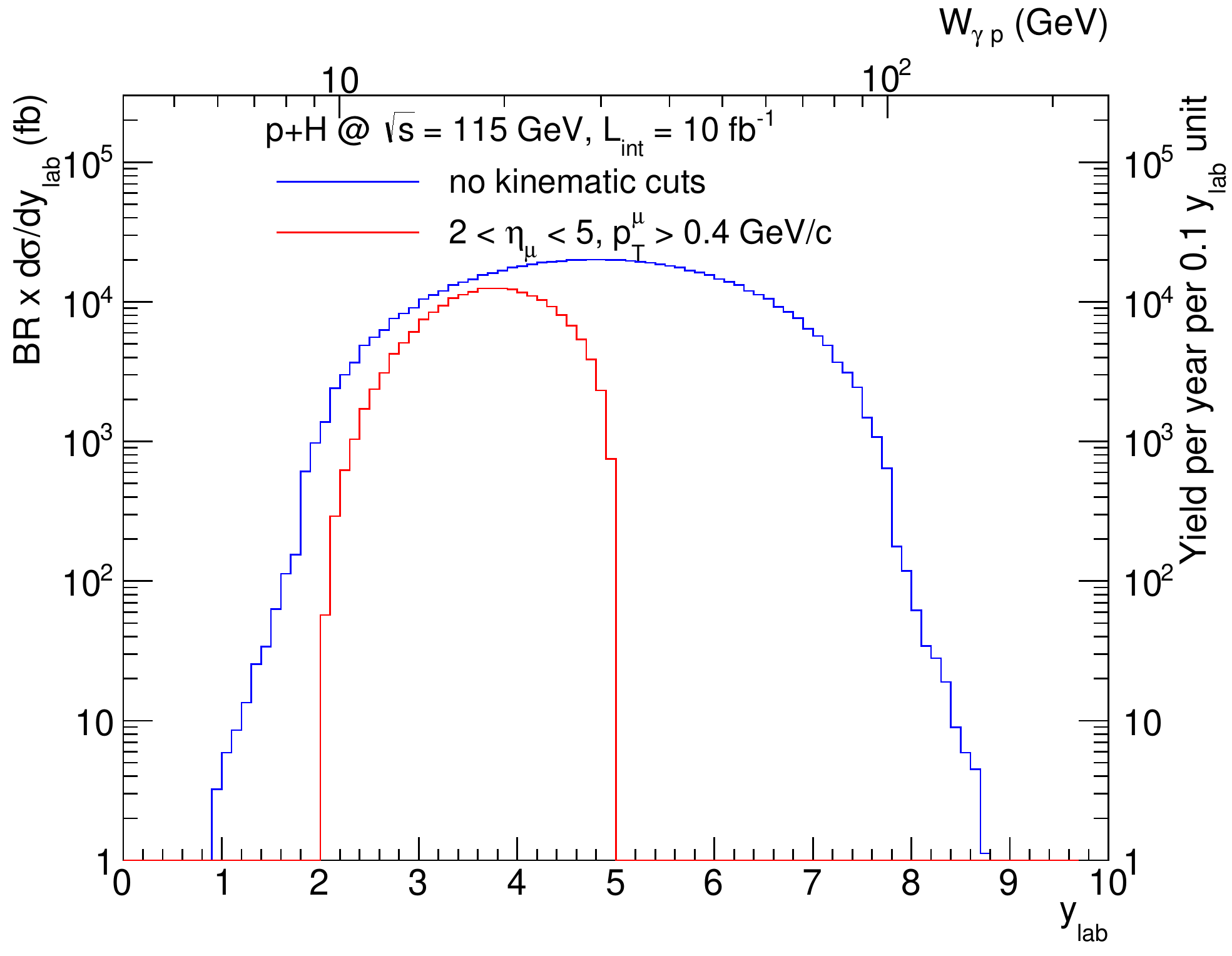}
 \includegraphics[width=7.9cm,clip]{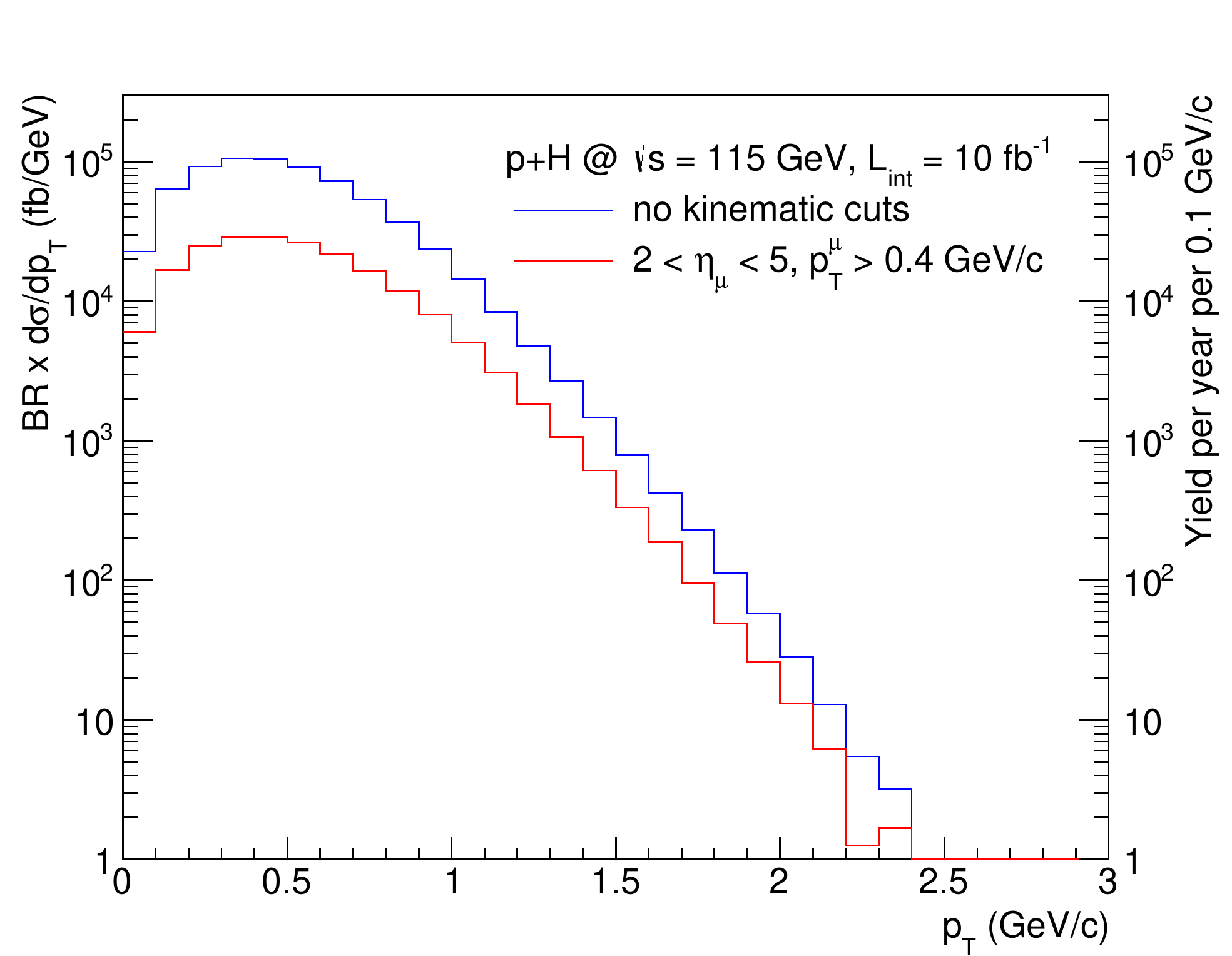}

\caption{$y_{\rm lab}$- (left)  and $p_{\rm T}$-differential (right) cross sections of the photoproduced $J/\psi$  in $p$H collisions generated with STARLIGHT MC. On the vertical axis is also shown the photoproduction $J/\psi$ yield per year per 0.1 $y_{\rm lab}$ unit (left), and per 0.1 GeV/c unit (right). On the left panel is indicated as well the photon-proton c.m energy (W$_{\gamma p}$). The blue curves have been produced without applying kinematics cut, while the red curves are produced by applying the $\eta$ and $p_{\rm T}$ cuts described in the text.}

\label{fig-1}       
\end{figure*}

\begin{figure*}
\centering
 \includegraphics[width=7.9cm,clip]{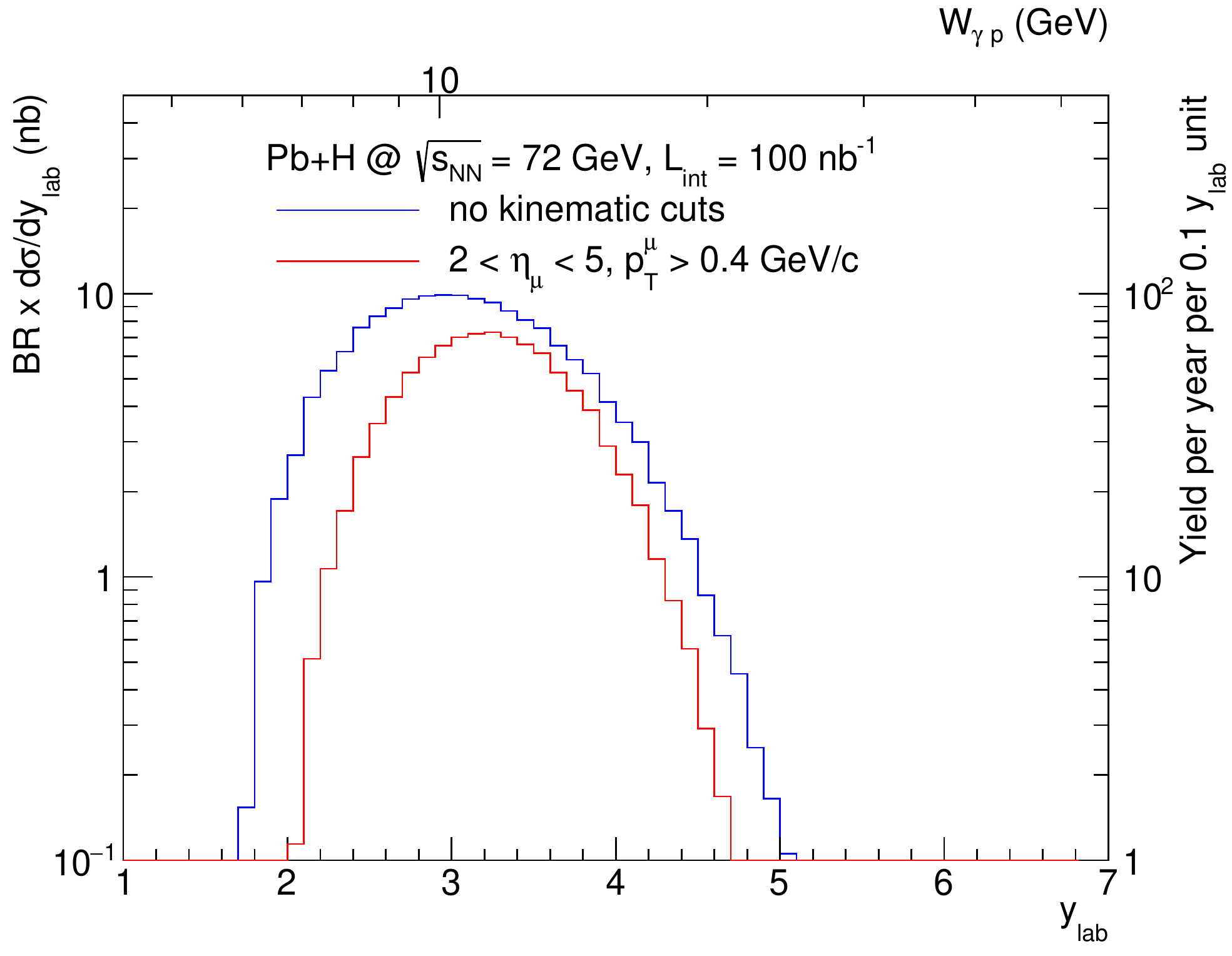}
 \includegraphics[width=7.9cm,clip]{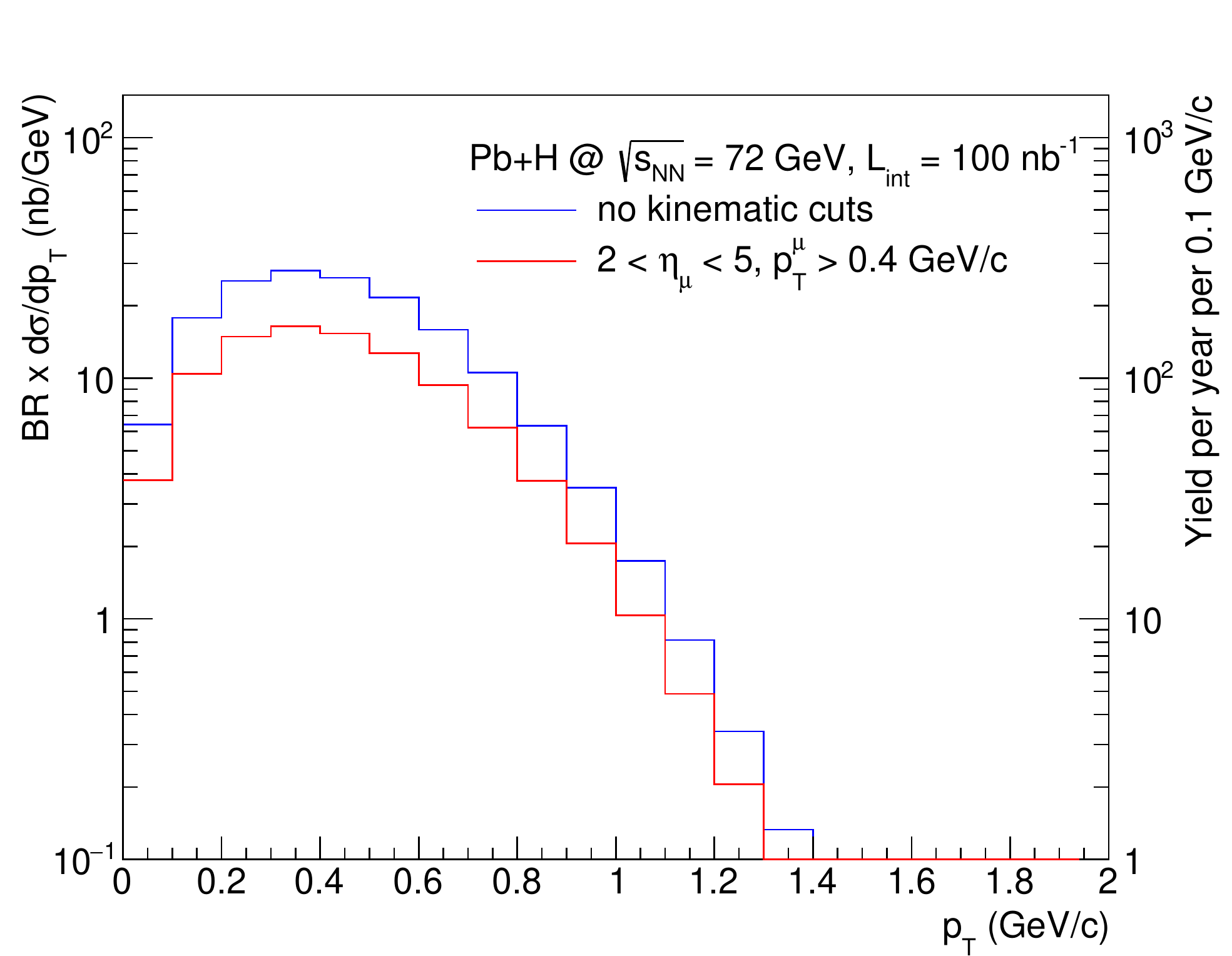}
\caption{Rapidity-differential (left) and $p_{\rm T}$-differential (right) cross sections of the photoproduced $J/\psi$ in the
laboratory frame in PbH collisions generated with STARLIGHT MC. On the vertical axis is also shown the photoproduction $J/\psi$ yield per year per 0.1 y$_{\rm lab}$ unit (left), and per 0.1 GeV/c unit (right). On the left panel is indicated as well the photon-proton c.m energy (W$_{\gamma p}$). The blue curves have been produced without applying kinematics cut, while the red curves are produced by applying the $\eta$ and $p_{\rm T}$ cuts described in the text.}
\label{fig-2}       
\end{figure*}

Let us now quickly summarise our feasibility study: 100~000 photoproduced $J/\psi$ have been generated in the dimuon decay channel, using STARLIGHT Monte Carlo (MC) generator \cite{Klein:1999qj,Baltz:2002pp,Klein:2003vd,Baltz:2009jk} in $p$H$^{\uparrow}$ ($\sqrt{s}$ = 115 GeV) and Pb+H$^{\uparrow}$ collisions ($\sqrt{s_{NN}}$ = 72 GeV). In $p$H$^{\uparrow}$ collisions, both protons can be photon emitters, while in Pb+H$^{\uparrow}$ only the Pb nuclei was considered as photon emitter (dominant contribution). The $J/\psi$ photoproduction cross sections given by STARLIGHT MC are summarised in Tab. \ref{Tab2}. In order to mimic an LHCb-like forward detector, the following kinematical cuts have been applied at the single muon level: 2 $< \eta_{\rm  lab}^{\mu} <$ 5  and $p_{\rm T}^{\mu} >$ 0.4 GeV/c. Fig. \ref{fig-1} shows the rapidity-differential (left) and $p_{\rm T}$-differential (right) cross sections of the photoproduced $J/\psi$ in the dimuon decay channel in $p$H$^{\uparrow}$ collisions, generated with STARLIGHT MC generator. The blue curves have been produced without applying any kinematical cuts, while the red curves are produced by applying the cuts: 2 $< \eta_{\rm  lab}^{\mu} <$ 5  and $p_{\rm T}^{\mu} >$ 0.4 GeV/c at the single muon level.  On the left panel is also indicated the photon-proton c.m energy (W$_{\gamma p}$), calculated as: 
\begin{equation}
W_{\gamma p} = \sqrt{M_{J/\psi}^{2} + M_{p}^{2} + 2 \times M_{p} \times M_{J/\psi} \times \cosh (\rm y_{\rm lab})},
\end{equation} 
with $M_{J/\psi}$ and $M_{p}$, being respectively the $J/\psi$ and proton masses, and $y_{\rm  lab}$ the $J/\psi$ rapidity in the laboratory frame.
On the vertical axis is shown as well the photoproduction $J/\psi$ yield per year per 0.1 $y_{\rm lab}$ unit (left) and per 0.1 GeV/c unit (right). An integrated yearly luminosity of $\cal{L_{\rm int}}$ = 10 fb$^{-1}$, corresponding to the storage cell solution,  has been assumed in $p$H$^{\uparrow}$ collisions in order to calculate the $J/\psi$ yearly photoproduction yield. About 200~000 photoproduced $J/\psi$ are expected to be detected in an LHCb-like acceptance per year with AFTER@LHC\footnote{The detector efficiency has been assumed to be 100$\%$. }. 
Similarly to Fig. \ref{fig-1}, Fig. \ref{fig-2} shows the rapidity-differential (left) and $p_{\rm T}$-differential (right) cross sections of the photoproduced $J/\psi$ in the dimuon decay channel in PbH$^{\uparrow}$ collisions, generated with STARLIGHT MC generator. The collection of an integrated luminosity of 100 nb$^{-1}$ per year is expected at AFTER@LHC with the storage cell option. This would result in about 1~000 photoproduced $J/\psi$ per year emitted in an LHCb-like acceptance. 

\begin{table}[!htpb]
\caption{Summary table of $J/\psi$ photoproduction cross sections from STARLIGHT MC generator.}
\resizebox{\textwidth}{!}{%
\begin{tabular}{|c|c|c|c|c|}
\hline
  collision type   & photon-emitter &  $\sigma^{tot}_{J/\psi}$ (pb)  & $\sigma_{J/\psi \rightarrow \mu^{+}\mu^{-}}$ (pb) &  $\sigma_{J/\psi \rightarrow \mu^{+}\mu^{-}}$ (2 $< \eta_{\rm lab}^{\mu} < $ 5, $p_{\rm T}^{\mu}$ > 0.4 GeV/c) (pb)  \\
  \hline
  $p$H$^{\uparrow}$  & proton (sum of 2 contributions) & 1.18$\times$10$^{3}$  & 70.10 & 20.64 \\
 \hline
   Pb+H$^{\uparrow}$ & Pb & 276.77$\times$10$^{3}$  & 16.50$\times$10$^{3}$ & 9.81$\times$10$^{3}$   \\ 
 \hline
\end{tabular} }
\label{Tab2}
\end{table}

In a forthcoming publication~\cite{forthcoming}, we will report on the evaluation of 
the uncertainty on the STSA, $A_{N}$, from the photoproduced-$J/\psi$ yields obtained with STARLIGHT MC 
and the expected modulation for $E_g$ following \cite{Koempel:2011rc}. Indeed, $A_{N}$, the amplitude of the spin-correlated azimuthal modulation of the produced particles, is defined as $A_{N} = \frac{1}{P}\frac{N^{\uparrow} - N^{\downarrow}}{N^{\uparrow} + N^{\downarrow}}$,
where $N^{\uparrow}$ ($N^{\downarrow}$) are the photoproduced-$J/\psi$ yields for an up (down) target-polarisation orientation, and $P$ is the effective polarisation of the target. The statistical uncertainty $u_{A_{N}}$ on $A_{N}$ can then  be derived using $
u_{A_{N}} = \frac{2}{P(N^{\uparrow} + N^{\downarrow})^{2}}\sqrt{N^{\downarrow2}u^{\uparrow2} + N^{\uparrow2}u^{\downarrow2}}$
with $u^\uparrow$ ($u\downarrow$) the relative uncertainties on the $J/\psi$ yields with up (down) polarisation orientation. Dividing the sample into two $J/\psi$ $p_{\rm T}$ ranges relevant for GPDs extraction: 0.4 $< p_{\rm T} <$ 0.6 GeV/c and 0.6 $< p_{\rm T} <$ 0.8 GeV/c, the expected statistical precision on $A_{N}$ is already expected to be better than 10$\%$ given the integrated yearly yield of 200~000 photoproduced $J/\psi$ in $pH^{\uparrow}$ collisions at AFTER@LHC. This will also allow for an extraction of A$_{N}$ as a function Feynman-$x$, $x_{\rm F}$ \footnote{$x_{\rm F} = 2 \times M_{J/\psi} \times \sinh ({y_{\rm cms}}/{\sqrt{s}})$,
with  $y_{cms}$ being the $J/\psi$ rapidity in the c.m. system frame and $\sqrt{s}$ the c.m energy}.




\section{Conclusion}

We have presented projections for $J/\psi$ photoproduction measurements in polarised $p$H$^{\uparrow}$ and Pb+H$^{\uparrow}$ collisions after one year of data taking at AFTER@LHC energies and assuming a storage cell technology coupled to an LHCb-like forward detector. In $p$H$^{\uparrow}$ collisions, a yearly yield of about 200~000 photoproduced $J/\psi$ is expected, allowing one to reach a very competitive statistical accuracy on the $A_{N}$ measurement differential in  $x_{\rm F}$. A non-zero asymmetry would be the first signature of a non-zero GPD $E_g$ for gluons.

\section*{Acknowledgements}

We thank D. Kikola, S. Klein, A. Metz, J. Nystrand and B. Trzeciak for useful discussions. This work is partly supported by the COPIN-IN2P3
Agreement, the grant of National Science Center, Poland, No. 2015/17/B/ST2/01838, by French-Polish scientific agreement POLONIUM, by the French P2IO Excellence Laboratory and by the French CNRS-IN2P3 (project TMD@NLO).


\newpage
\bibliographystyle{utphys_cp}
\bibliography{cernrepexa}


%
%
%
%
%

\end{document}